\documentclass[letter]{aa} 

\newcommand{\teff}{$T_{\rm eff}$}
\newcommand{\omcen}{$\omega$~Cen}
\newcommand{\mh}{$[M/H]$}

\usepackage{natbib}
\usepackage{graphicx}
\usepackage{booktabs}
\usepackage{txfonts}
\usepackage{hyperref}
\begin{document}

   \title{A stellar census in globular clusters with MUSE.}

\subtitle{A new perspective on the multiple main sequences of $\omega$ Centauri\thanks{ Based on observations collected at the European Organisation for Astronomical Research in the Southern Hemisphere, Chile (Program IDs 094.D-0142(B), 095.D-0629(A), 096.D-0175(A), 097.D-0295(A), 098.D-0148(A), 099.D-0019(A), 0100.D-0161(A), 0101.D-0268(A), 0102.D-0270(A), 0103.D-0204(A), and 0104.D-0257(B))}$^,$\thanks{Table B.1 is only available at the CDS via anonymous ftp to
cdsarc.u-strasbg.fr (XXXX) or via http://cdsarc.u-strasbg.fr/viz-bin/cat/J/A+A/XXX/Lzzz}}

   \author{M. Latour\inst{1},
          A. Calamida\inst{2},
          T.-O. Husser\inst{1}, S. Kamann\inst{3}, S. Dreizler\inst{1}, \and
           J. Brinchmann\inst{4,5}
          }

   \institute{Institute for Astrophysics, Georg-August-University, Friedrich-Hund-Platz 1, 37077 G\"{o}ttingen, Germany, \email{marilyn.latour@uni-goettingen.de}
    \and Space Telescope Science Institute, 3700 San Martin Drive, 
Baltimore, MD 21218, USA
    \and
    Astrophysics Research Institute, Liverpool John Moores University, 146 Brownlow Hill, Liverpool L3 5RF, United Kingdom 
    \and
    Instituto de Astrof\'isica e Ci\^encias do Espaço, Universidade do Porto, CAUP, Rua das Estrelas, PT4150-762 Porto, Portugal
    \and
    Leiden Observatory, Leiden University, PO Box 9513, 2300 RA Leiden, The Netherlands
             }

   \date{Received 15 June 2021 / Accepted 3 September 2021}

 
  \abstract
   {\omcen\ is a rare example of a globular cluster where the iron abundance of the stars spans more than one order of magnitude. Many spectroscopic investigations of its red-giant and sub-giant branches have revealed multiple peaks in the iron abundance distribution. The metallicity distribution of main sequence (MS) stars is not well characterized yet due to the faintness of the stars and lack of data. So far, almost all studies of MS stars have been based on photometric measurements.}  
   {Our goal is to investigate the metallicity distribution of a statistically significant sample of MS stars in \omcen. In particular, we aim to revisit the metallicity difference between the red and blue MS of the cluster.}
   {We used
   MUSE spectra obtained for the central region of \omcen\ to derive metallicities for $\approx$4200 MS stars.}
   {We find that blue MS stars are, on average, $\approx$0.1 dex more metal-rich than their red counterparts. On the basis of this new estimate, we find that the two sequences can be fit on the Hubble Space Telescope color-magnitude diagram with two isochrones having the same global metallicity and age, but 
   a higher helium abundance for the blue MS, that is
   $\Delta Y \lesssim$ 0.1. Furthermore, we determined the average metallicity of the five main populations along \omcen~MS and these estimates are consistent with expectations from previous photometric studies.}
   {}
\authorrunning{M. Latour et al.} 
\titlerunning{Metallicities on the multiple main sequences of \omcen}
   \maketitle
%

\section{Introduction}

\omcen\ is the most massive Galactic globular cluster (GGC), with a mass of 3.5 $\times$ 10$^6$ $M_\odot$ \citep{baumgardt2018}.
The iron abundance of its stars spans more than one order of magnitude $ -2.2 \lesssim [Fe/H] \lesssim -0.6 $ and light-element abundance variations 
are observed within groups of stars along this metallicity range  \citep{norris_dacosta1995, suntzeff1996, calamida2009, johnson2010, marino2011, husser2020}.
All these studies find a metallicity distribution for red-giant branch (RGB) stars with multiple main peaks around $[Fe/H] \approx$ -1.7, -1.4, -1.0, and -0.8. The large range of iron abundances observed among cluster RGB stars was also detected among sub-giant branch (SGB) and main sequence turn-off (MSTO) stars \citep{kayser2006, hilker2004, sollima2005b, stanford2006, villanova2007, pancino2011a}. 

The chemical picture of \omcen's stars is so complex that there currently are different hypotheses on its formation.
The two main hypotheses are that \omcen\ is the nucleus or the nuclear star cluster of a dwarf galaxy accreted by the Milky Way or the result of the merger of two or more clusters \citep{norris1997,pancino2000, bekki2006,massari2019, ibata2019, alfaro2019, calamida2020, pfeffer2021}. In particular, the merger of smaller clusters or a cluster and a nuclear stars cluster could have happened in a dwarf galaxy, where these encounters are more frequent than in the Galactic halo due to the lower velocity dispersion, and the system could have been later accreted by the Galaxy  \citep{amaroseoane2013, bekkitsu2016,gavagnin2016,pasquato2016}.

A bifurcation of the main sequence (MS) of \omcen\ into two main components, the so-called blue MS (bMS) and red MS 
(rMS), was first revealed by Hubble Space Telescope (HST) photometry \citep{anderson2002, bedin2004}. Although the discrete nature of the cluster's RGB was known from previous photometric investigations \citep{ lee1999, pancino2000},
this was the first time that such a bimodal distribution of stars was observed along the MS of a GGC. Follow up spectroscopy of a handful of MS stars by \citet[hereinafter PI05]{piotto2005} indicated that the stars belonging to the bMS had a higher metallicity ($+0.3\pm0.2$ dex) compared to stars found on the rMS. This reversal of the expected color progression with metallicity was explained in terms of different helium abundances, that is, the bMS stars would contain a much higher helium proportion than the stars on the rMS ($\Delta Y \approx0.12-0.15$; \citealt{Norris2004}, PI05, \citealt{king2012}).

Numerous studies tried to characterize the properties of the bimodal MS, such as the radial distribution, age, and chemical composition \citep{sollima2007c, king2012, calamida2017, calamida2020, tailo2016, milone2017c}.
All these studies were based on photometric investigations. 
In recent years, more subpopulations were identified along the MS from the initial two: a high-metallicity component called MSa, which can distinctly be followed through the sub-giant branch and onto the RGB (SGB-a, RGB-a), and \citet[hereinafter BE17]{bellini2017c} separated 15 subpopulations by using pseudo-color diagrams based on a combination of the $F275W$, $F336W$, and $F438W$ WFC3/HST filters. These subpopulations are divided into the following five main components in the 
color-magnitude diagrams (CMDs) of \omcen: the rMS, bMS, MSa, plus the newly identified MSd and MSe.
The chemical properties of these populations were qualitatively assessed thanks to the use of magnitudes in multiple filters ranging from the near-UV to the near-infrared \citep{bellini2017a}. While the effect of $CNO$ variations leaves a rather clear signature in the WFC3/HST F336W and F438W filters, the effects of helium, and especially iron, are more subtle (see Sect. 5 of BE17). Although very insightful, the information that can be gained from photometry is limited and this is where spectroscopy proves essential to gain a clearer picture.
While a handful of spectroscopic surveys of RGB, SGB, and MSTO stars in \omcen\ have been carried out in the past, no spectroscopic studies of MS stars have been done since PI05. The faintness of the MS stars combined with the highly crowded field make such observations extremely challenging and time-consuming. Indeed, PI05 observed only 34 MS stars using the multi-object spectrograph FLAMES on the Very Large Array Telescope (VLT) for a total exposure time of 12h. 
The spectra of the bMS and rMS stars were then co-added, resulting in signal-to-noise ratios (S/Ns) of $\approx$30 at a spectral resolution $R=6400$ (3960-4560 \AA). The authors analyzed the two spectra that were considered representative of the bMS and rMS stars.
   
In this Letter, we present new spectroscopic observations
of MS stars in \omcen\ after more than fifteen years. Our goal is to revisit the properties of the red and blue MS based on spectroscopic evidence and look at the metallicity distribution of the MS subpopulations identified by BE17. To achieve this, we use MUSE observations of \omcen\ obtained as part of the GTO time dedicated to GGCs (PIs: S. Dreizler, S. Kamann). 
   
\section{MUSE observations and spectral analysis}\label{sec:obs}

We use the MUSE GTO observations of \omcen\ obtained in wide field mode (1$\arcmin \times$ 1$\arcmin$) between April 2014 and March 2019. 
The spectra cover the 4750$-$9350 \AA\ range with an average spectral resolution of $\sim$2.5 \AA\ \textbf{($R \sim3000$)}, although this varies slightly across the wavelength range \citep{husser2016}.
A general description of the data reduction and spectral analysis is presented in \citet{husser2016} and \citet{kamann2018}. 
Additional information on the MUSE data used for this work is included in Appendix A.
In the following, we briefly provide some details specific to the analysis of the \omcen\ MS spectra.

Because each of the seven MUSE fields was observed multiple times, the individual spectra of each star were co-added after being corrected for their measured radial velocities (RVs). 
The combined spectra were then fitted using the spectrum fitting framework \textsc{spexxy}\footnote{\url{https://github.com/thusser/spexxy}} and the G\"{o}ttingen spectral library of PHOENIX models \citep{husser2013}. The best fit provides \teff\ and \mh, while the log $g$ was fixed to the value provided by the isochrones and the $\alpha$-enhancement was kept constant at $[\alpha/Fe]$=0.3. 
For \omcen, up to three sets of photometry are available per star: from \citet{sarajedini2007}, \citet{bellini2017a}, and from the archival ACS observation for Field \#7 (see Sect. 4 of \citealt{kamann2018}). For each of these photometric sets, a best fitting isochrone from the \textsc{parsec} database \citep{marigo17} was selected to simultaneously match the position of the MS, SGB, and RGB. The log $g$ of a star was then determined by finding the nearest point on the isochrone and the average log~$g$ was adopted when the star is present in more than one photometric catalog.
We note that this approach differs from previous spectroscopic studies of brighter stars in GCs. In literature work, \teff\ is usually obtained from empirical color-temperature relations (e.g., \citealt{alonso99}) and $[Fe/H]$ is obtained from the direct fit of Fe lines in high-resolution spectra. We fit the whole spectrum of our stars where the strongest metallic features are the Mg and Ca triplets. For this reason, we keep the \mh\ nomenclature to refer to our metallicities, as it is closer to a global metallicity and is sensitive to star-to-star variations in Mg and Ca abundances.

We paid particular attention to the uncertainties of the metallicity measurements and performed a calibration following a procedure similar to the one used for RVs in \citet{kamann2016}. 
For each star, we obtained a correction factor by which we multiplied the formal uncertainty returned by the fitting procedure.

   \begin{figure}
   \centering
   \includegraphics[width=0.95\columnwidth]{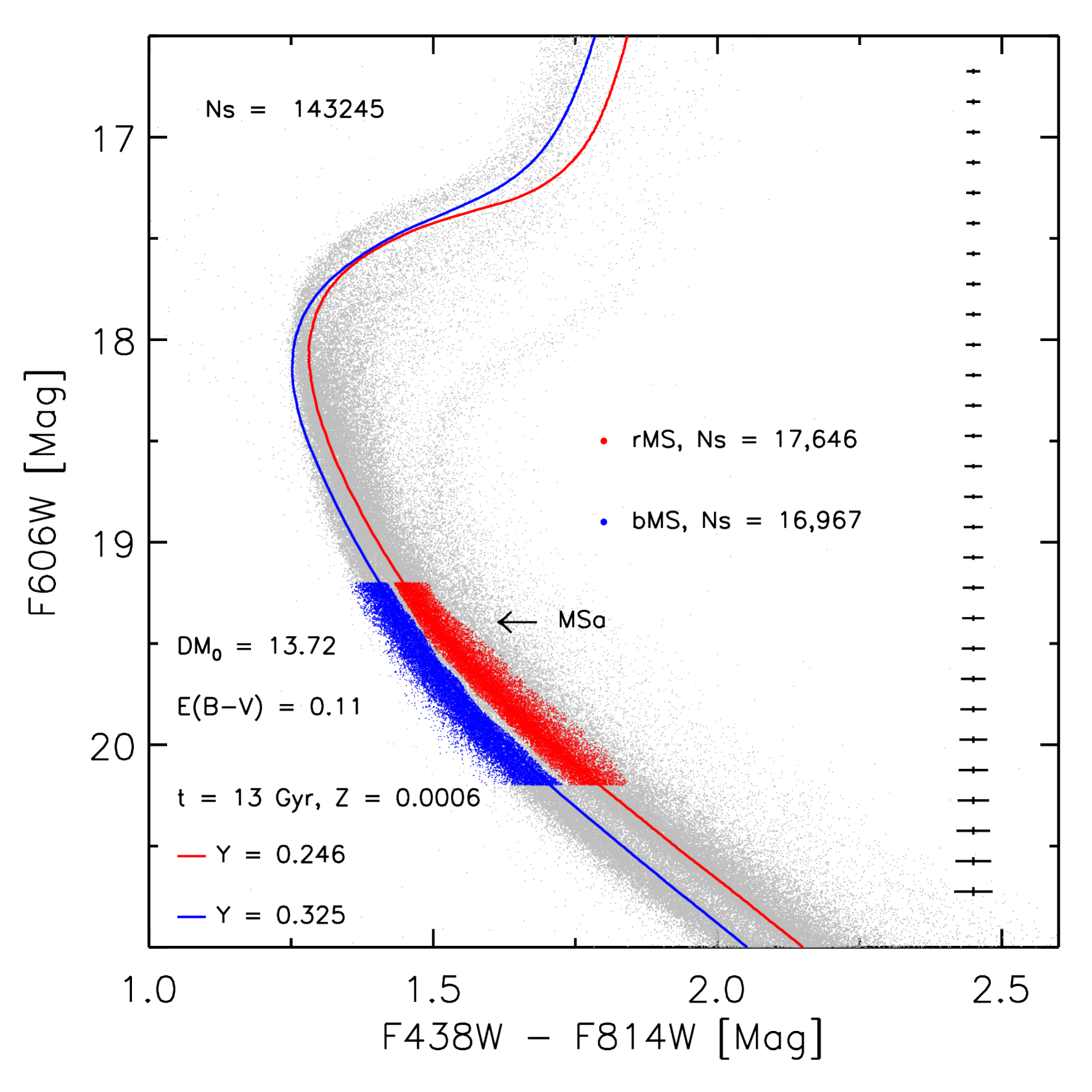}
   \caption{$F606W, F438W-F814W$ CMD from the photometric
catalog of  \cite{bellini2017a}. The selected samples of blue and red MS stars are marked with blue and red dots, respectively. Two isochrones from the BASTI database with different helium content and the same metallicity and age are overplotted.}
    \label{fig_cmd}
    \end{figure}

\section{Selection of red and blue main sequence stars}\label{sec:selection}
In order to select our blue and red MS star samples, we retrieved the WFC3/HST photometric catalog of \citet{bellini2017a}. We first used the $F606W,\ F438W - F814W$ CMD since it is sensitive to metallicity changes (or helium variations) and less to light-element abundance variations \citep{calamida2017}. At this stage, we were only interested in selecting the two main groups along \omcen's MS. 
We then included all stars in the magnitude range 19.2 $< F606W <$ 20.2 along the blue and red MS, making sure that the two selected boxes were not overlapping in the $F438W - F814W$ color. Stars on the MSa, lying on the reddest part of the MS in this CMD, were left out of the selection (see Fig.~\ref{fig_cmd}).
We selected stars in this magnitude range as a trade-off between obtaining a better separation of the bMS and rMS stellar populations and having a reasonable $S/N$ from the MUSE spectra. The two MS samples also have a very similar completeness level since they are spanning the same magnitude range.
The selected bMS and rMS samples were also plotted on the $F606W,\ F336W-F438W$ and $F606W,\ F275W-F438W$ CMDs (these two colors are very sensitive to light-element abundance variations) to eliminate cross-contamination between the two groups. 
The two samples are shown in Fig.~\ref{fig_cmd} as red dots (rMS) and blue dots (bMS) and include 17\,646 and 16\,967 stars, respectively.

\section{Metallicity distributions of MS stars}

\subsection{The rMS and bMS}

   \begin{figure}
   \centering
   \includegraphics[width=0.9\columnwidth]{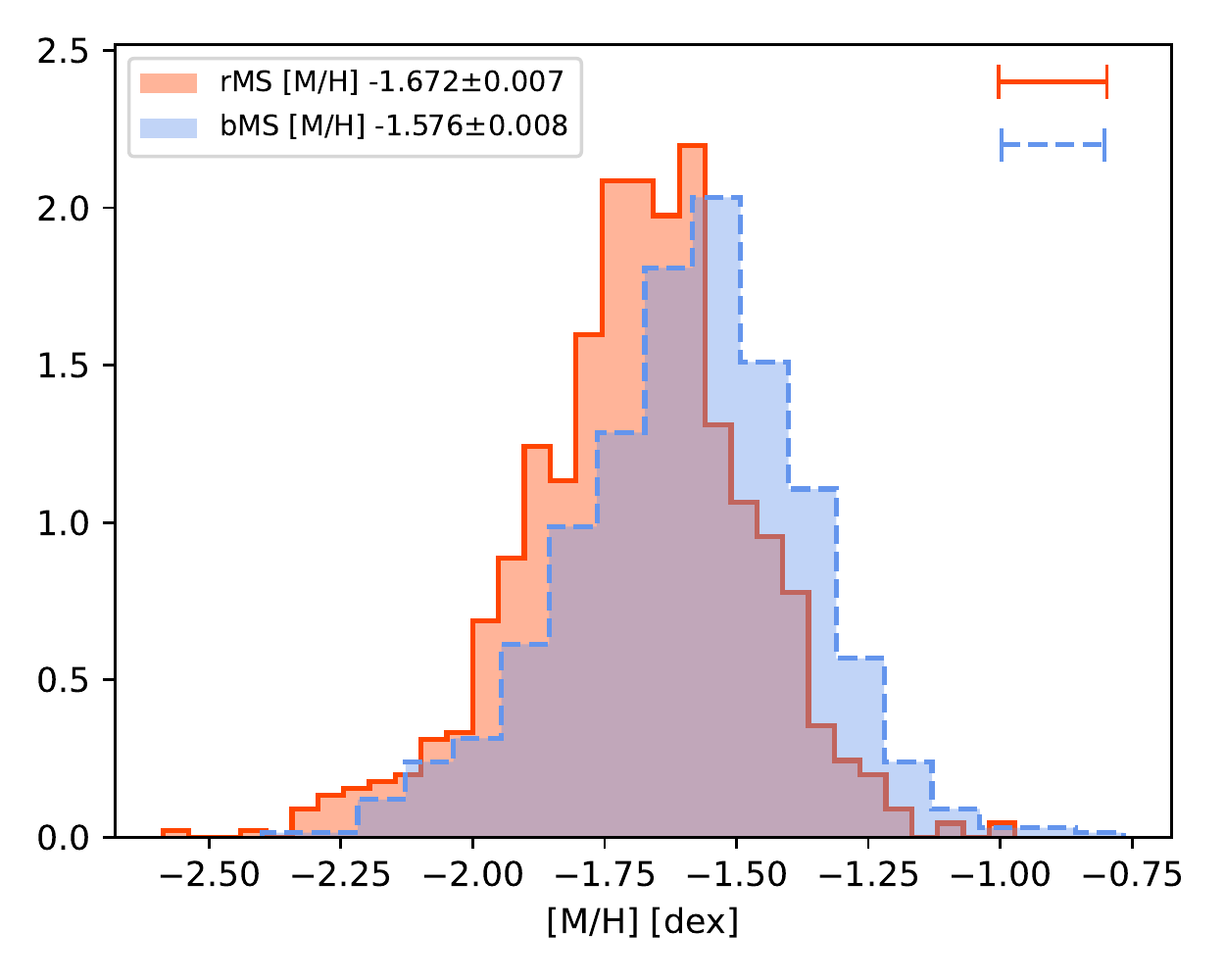}
      \caption{Normalized metallicity distribution for the stars with S/N $>$ 20 belonging to the rMS (solid line, 920 stars) and the bMS (dashed line, 737 stars). The mean [M/H] value of each distribution is indicated in the legend. The horizontal bars in the top right corner show the median value ($\pm$0.1 dex) of the corrected [M/H] uncertainties.}
         \label{hist_bMS_rMS}
   \end{figure}

%
\begin{table}
\small
\caption{Properties of the rMS and bMS stars}
\label{table_bMS_rMS}      
\centering                    
\begin{tabular}{c c l c c}        
\toprule\toprule              
MS &  $S/N$ & \# of stars & Mean $[M/H]$ & $\sigma$ $[M/H]$ \\   
\midrule      
rMS & $>$ 15 & 1629 & $-$1.669 $\pm$ 0.005 & 0.179 $\pm$ 0.005\\
rMS & $>$ 20 & 920 & $-$1.672 $\pm$ 0.007 & 0.172 $\pm$ 0.006 \\
bMS & $>$ 15 & 1358 & $-$1.573 $\pm$ 0.006 & 0.202 $\pm$ 0.005 \\
bMS & $>$ 20 & 737 & $-$1.576 $\pm$ 0.008 & 0.192 $\pm$ 0.006 \\
\bottomrule                                
\end{tabular}
\end{table}
%
We matched the targets selected in Sect.~\ref{sec:selection} with our MUSE database to retrieve the combined spectra of the stars that are in the MUSE fields 
and fitted them as described in Sect.~\ref{sec:obs}.
We applied a few additional criteria to define our final sample of red and blue MS stars:
we computed results for two different $S/N$ cuts, 15 and 20, and we then removed stars with a membership probability below 50\%. 
This is based on the combination of a given star's RV and $[M/H]$ (see \citealt{kamann2016} for more details); in the case of \omcen, the high RV of the cluster (232 km\,s$^{-1}$) is the main discriminating factor. Finally, we eliminated stars for which the standard deviation of the log $g$ obtained from the different isochrones is larger than 0.1 dex; these stars have an inconsistent magnitude in one of the catalogs, giving them a different position in the CMDs. 
The last two criteria exclude only a small amount of stars. The number of stars left in our samples is shown in Table \ref{table_bMS_rMS}, including  several hundred stars in all cases. 

The metallicity distributions of rMS and bMS stars with $S/N$~$>$~20 are illustrated in Fig.~\ref{hist_bMS_rMS}. The median $S/N$ of these samples is 25. To estimate the average \mh, the dispersion, and their uncertainty, we used the Markov chain Monte Carlo (MCMC) ensemble sampler developed by \citet{emcee2013}. We used a simple Gaussian model to reproduce the observed distributions and retrieve the most likely mean ($\mu$) and standard deviation ($\sigma)$ values. The results are listed in Table \ref{table_bMS_rMS} where the uncertainties indicate the 16$^{\rm th}$ and 84$^{\rm th}$ percentile of the posterior probability distributions (equivalent to 1$\sigma$). We find that the bMS stars are slightly more metal-rich than the rMS stars; the difference of +0.096 $\pm$ 0.011 dex has a significance of 8$\sigma$. This is at the lower limit of the +0.3$\pm$0.2 dex range obtained by PI05. Our result, based on a statistically significant sample of MS stars ($>$1000 vs 34), clearly rules out a 0.3 dex difference between \omcen's bMS and rMS. This value and the consequence of bluer MS stars being highly helium-enhanced compared to redder ones were assumed to interpret the CMDs and the origin of multiple populations in \omcen\ and many other GGCs in 
subsequent years (e.g., \citealt{milone2017c}). However, it is worth pointing out that the two sequences used in PI05 are likely not equivalent to ours, nor to those of  BE17, because PI05 have a ratio of bMS to rMS stars $\approx$1:3 while our numbers, and those of BE17, are closer to a 1:1 ratio.
We also find that the average metallicity of both MSs is not significantly affected by the $S/N$ cut. 
Because the MCMC procedure takes the calibrated \mh\ uncertainties  into account, the value of $\sigma$ is representative of the intrinsic dispersion of metallicities within each population. The values obtained for our samples suggest the presence of a small metallicity spread 
and it is in fact plausible that the two populations are not strictly mono-metallic. 
After all, \citet{villanova2014} found that it was not the case for the multiple SGBs of the cluster and \citet{tailo2016} reproduced, via stellar population synthesis, the bMS and rMS by using a mixture of metal-poor and metal-intermediate stars having different helium content.

\subsection{The five main sequences of Bellini et al.}

\begin{table}
\small
\caption{Properties of MSs identified in \citet{bellini2017c}}
\label{table_5MS}      
\centering                    
\begin{tabular}{l c l c c}        
\toprule\toprule
MS &  $S/N$ & \# of stars & Mean $[M/H]$ & $\sigma$ $[M/H]$ \\   
\midrule                      
rMS & $>$ 20 & 464 & $-$1.661 $\pm$ 0.009 & 0.160 $\pm$ 0.008 \\
rMS1 & $>$ 15 & 248 & $-$1.663 $\pm$ 0.013 & 0.174 $\pm$ 0.012 \\
rMS2 & $>$ 15 & 302 & $-$1.641 $\pm$ 0.012 & 0.160 $\pm$ 0.010 \\
rMS3 & $>$ 15 & 329 & $-$1.667 $\pm$ 0.012 & 0.172 $\pm$ 0.011 \\
bMS & $>$ 20 & 341 & $-$1.557 $\pm$ 0.011 & 0.186 $\pm$ 0.009 \\
bMS1 & $>$ 15 & 313 & $-$1.663 $\pm$ 0.013 & 0.187 $\pm$ 0.011 \\
bMS2 & $>$ 15 & 169 & $-$1.517 $\pm$ 0.016 & 0.170 $\pm$ 0.014 \\
bMS3 & $>$ 15 & 231 & $-$1.413 $\pm$ 0.012 & 0.148 $\pm$ 0.011 \\
MSe & $>$ 20 & 292 & $-$1.504 $\pm$ 0.010 & 0.150 $\pm$ 0.009\\
MSd & $>$ 20 & 78 & $-$1.228 $\pm$ 0.017 & 0.122 $\pm$ 0.014 \\
MSa & $>$ 20 & 250 & $-$0.964 $\pm$ 0.009 & 0.138 $\pm$ 0.008\\
\bottomrule                          
\end{tabular}
\end{table}
%

   \begin{figure}
   \centering
   \includegraphics[width=0.9\columnwidth]{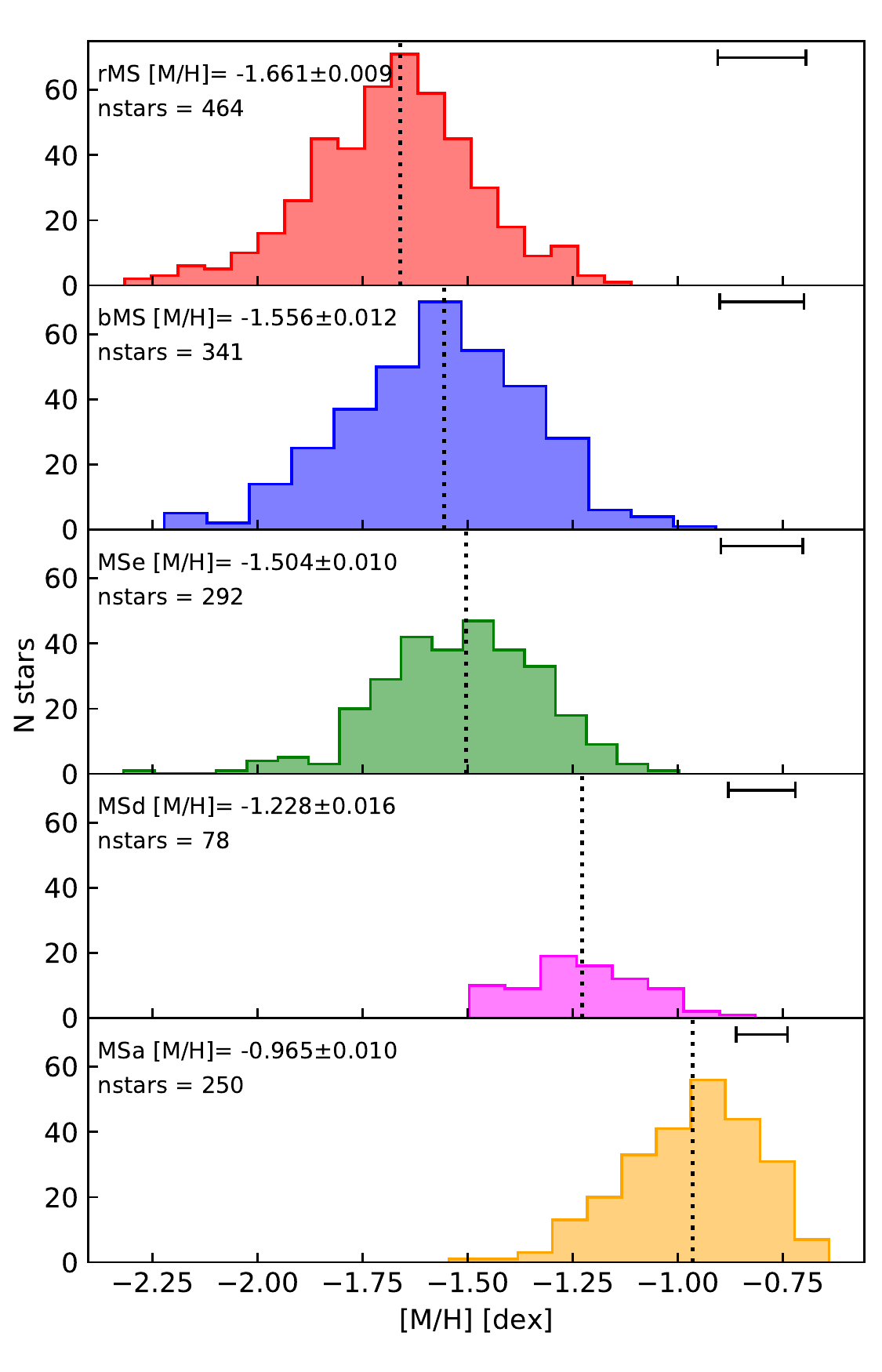}
      \caption{Normalized metallicity distributions for the stars with S/N $>$ 20 belonging to the five MSs defined by BE17. The mean [M/H] value (dotted line) of the distributions and the number of stars included are indicated in each panel. The horizontal bars in the top right corner show the median value of the corrected [M/H] uncertainties.}
         \label{hist_5MS}
   \end{figure}

In addition to the identification of 15 subpopulations across the MS of \omcen, BE17 made qualitative predictions about the chemical composition (Fe, He, and N) of these populations based on their photometric properties. Since the population tags are available, we compared our metallicities with their predictions of iron abundances. We selected our spectroscopic samples for the five main components (MSa, bMS, rMS, MSd, and MSe) following the same criteria as in the previous subsection. We list their properties in Table \ref{table_5MS}, and show their metallicity distributions in Fig.~\ref{hist_5MS}. In order of increasing average metallicity, we find the following sequence: rMS - bMS - MSe - MSd - MSa. We obtain a difference of $+$0.104 $\pm$ 0.014 dex in metallicity between the rMS and bMS, which is in excellent agreement with the value obtained from our own selection. 
The metallicity progression between the five MSs is in line with the expectations of BE17. The authors attributed a similar Fe enrichment for the bMS and the MSe stars, but we find a $3\sigma$ evidence for the MSe stars to be more metal-rich than the bMS stars. These two populations are the ones with the closest mean [M/H]. The dispersion ($\sigma$) obtained for each population suggests the presence of a small ($<$0.2 dex) intrinsic spread in metallicity. 

Finally, we investigate whether these main populations are mono-metallic or not. Each of them is subdivided into two to four subpopulations in BE17 depending on the distribution of their stars in the chromosome map (see Appendix~\ref{app_c}). 
Here, we focus on the divisions of the rMS and bMS, while a discussion on the metallicities of all MSs is presented Appendix~\ref{app_c}. We used spectra with $S/N$ $>$15 in order to increase the number of stars in each sample.
The results are listed in Table \ref{table_5MS} and the distributions are shown in Fig.~\ref{rMS_bMS_split}. We find a different behavior between the rMS and bMS subpopulations: while the three rMSs have very similar metallicity distributions and average \mh, the three bMSs clearly have different metallicities. The metallicity increases from bMS1 to bMS3 with these two subpopulations having a difference of 0.25$\pm$0.018 dex. In fact, the bMS1, with a mean \mh$=-1.66$, is as metal-poor as the rMS.

\subsection{Comparison with spectroscopic studies in literature}
We compared our metallicity distribution for a selected sample of MS stars with the most recent high-resolution spectroscopy for \omcen\ RGB stars from \citet[hereinafter JO20]{johnson2020}. We also added spectra for 855 RGBs presented in \citet[hereinafter JO10]{johnson2010}, since the study of JO20 is biased toward metal-poor stars. We combined the two samples and obtained a list of 1250 RGBs with a $[Fe/H]$ measurement.
We compared MUSE metallicities, $[M/H]$, directly to JO20+JO10 $[Fe/H]$ values since the synthetic spectra used for the spectral fits are $\alpha$-enhanced ($[\alpha/Fe]$ = 0.3).

\begin{figure}
   \centering
   \includegraphics[width=0.9\columnwidth]{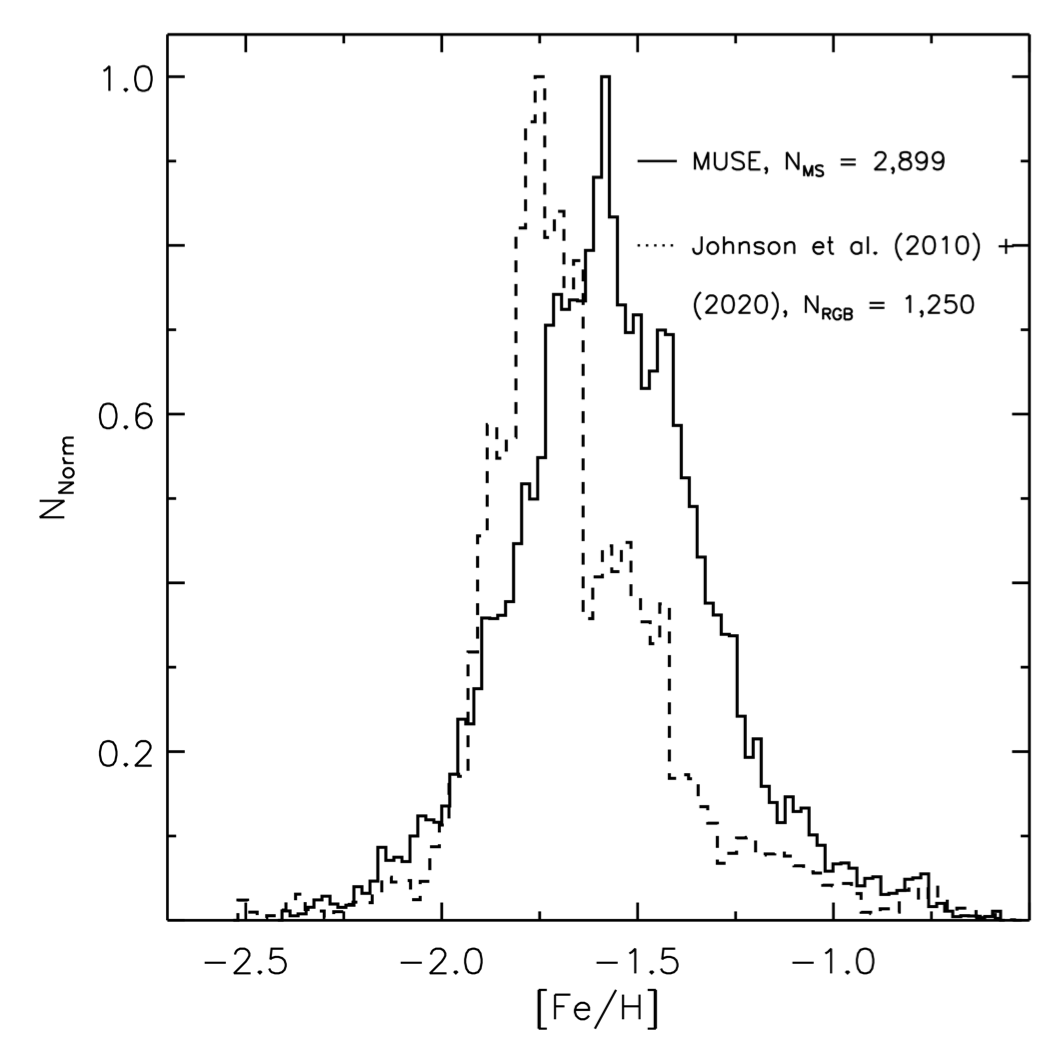}
      \caption{Comparison between the metallicities obtained from the MUSE spectra of MS stars and the [Fe/H] measured in RGB stars \citep{johnson2010,johnson2020} 
              }
         \label{hist_johnson}
   \end{figure}

Fig.~\ref{hist_johnson} shows the normalized MUSE metallicity distribution for 2899 MS stars in the magnitude range 19.2 $\lesssim F606W \lesssim$ 20.2 compared to the normalized and combined JO20+JO10 metallicity distribution for 1250 RGB stars. The shapes of the distributions are in good agreement: they cover the same wide metallicity range from $[Fe/H] \approx$-2.5 up to $\approx$-0.5 with a fast rise at low metallicity and a tail toward higher values. This qualitative comparison shows that the main peak is present at $[Fe/H]\approx$ -1.75 and a secondary one at $\approx$-1.5 for JO20+JO10. Two such peaks are also present in the MUSE sample, albeit at slightly more metal-rich positions; $[M/H]\approx$ -1.6 and $\approx$-1.4. 
Both distributions also show a shoulder at metallicities lower than the main peak.
The most metal-rich stars, $[Fe/H]\gtrsim$ -1.0, in the MUSE distribution are $\approx$ 3\% of the sample, which is in good agreement with previous photometric and spectroscopic studies that identified the most metal-rich population of \omcen, the RGBa, corresponding to the MSa, to include $\approx$ 3\% of the cluster stars \citep{pancino2003, castellani2007, bellini2017c}.
The systematic difference in metallicity of $\approx$ 0.15 dex between the two distributions is not unexpected. Systematic offsets with the iron abundances of JO10 have been reported in the past for RGB stars ($-$0.13 dex by \citealt{marino2011} and $+$0.1 dex by \citealt{meszaros2021}). These differences were partly explained by the use of different \teff. In our case, it must be kept in mind that we are looking at different stars (MS versus RGB) and that we used a different technique to derive \teff\ and \mh. In addition, knowing that the Mg abundances are varying by $\approx$1 dex in the cluster's RGB stars \citep{meszaros2021}, and assuming some variation in the MS stars as well, we can expect an additional broadening of our MS \mh\ distribution when compared to the RGB $[Fe/H]$ one.

\subsection{Comparison with models}
On the basis of the new spectroscopic metallicities derived for the bMS and rMS, a set of BASTI isochrones was used to reproduce \omcen\ MS. In particular, we adopted two $\alpha$-enhanced isochrones with the same age, t $=$ 13 Gyr, same global metallicity, Z$=$0.0006 ($[M/H] \approx$ -1.5), and different helium abundances, canonical, Y $=$ 0.246 and Y $\approx$ 0.32 \citep{pietrinferni06}. 
A distance modulus of $DM_0 =$ 13.72 \citep{braga2016} and reddening $E(B-V)=$ 0.11 \citep{calamida2005} were assumed. This reddening value was converted into extinctions in WFC3/HST filters by using the \citet{cardelli89} reddening law and the available WFC3 filter throughputs\footnote{\url{https://www.stsci.edu/hst/instrumentation/wfc3/performance/throughputs}}.
The two isochrones are overplotted to the 
CMD in Fig.~\ref{fig_cmd}: the canonical helium isochrone (red line) fits the rMS of \omcen\ very well from the base of the RGB down to the lower MS, and the helium-enhanced one (blue) for the same metallicity and age nicely fits the bMS. We find that, within the uncertainties, the blue and red MS in \omcen\ can be reproduced by models with similar average metallicities and ages, but with different helium contents, that is, $\Delta Y \approx$ 0.08, which is significantly less than the $\Delta Y =$ 0.12-0.15 derived by PI05.

\section{Conclusion}
We used MUSE spectra of $\approx$4200 MS stars in \omcen\ to derive their global metallicity \mh\ and examined the metallicity distribution and average value of the different MSs that have been identified in the cluster. This is the first time since PI05 that metallicities of MS stars in \omcen\ are obtained from spectroscopic data. 
We find the bMS stars to be slightly more metal-rich than the rMS stars, on average, with a difference of $+$0.10 $\pm$ 0.015 dex. This is at the lower limit of the range obtained by PI05. However, when looking at the bMS and rMS subdivisions identified by BE17, we discovered a more complex behavior: the three rMS subpopulations share a similar \mh\, while 
the three bMS subpopulations have metallicities that differ by $\approx$0.1 dex each, with the bluest sequence, bMS1, having the same metallicity as the rMS.
We were able to reproduce the position of rMS and bMS stars in the CMD with isochrones of the same age and metallicity ($Z$=0.0006) and a helium difference $\approx$ 0.08, which is a smaller helium enhancement than claimed thus far for \omcen\ bMS ($\Delta Y \approx$ 0.12-0.15, \citealt{piotto2005, Norris2004, king2012}).
We derived average metallicities for all MSs identified in BE17 and our results for the five main populations confirm the qualitative expectation for iron abundances made by the authors. 

\begin{acknowledgements}
 We thank G. Bono and N. Neumayer for useful comments and S. Cassisi, M. Salaris,  and A. Pietrinferni for providing us appropriate isochrones to fit the presented CMD.
 We acknowledge funding from the Deutsche Forschungsgemeinschaft (grant DR 281/35-1 and KA 4537/2-1) and from the German Ministry for Education and Science (BMBF Verbundforschung) through grants 05A14MGA, 05A17MGA, 05A14BAC, and 05A17BAA.
 SK gratefully acknowledges funding from UKRI in the form of a Future Leaders Fellowship (grant no. MR/T022868/1).
 JB acknowledges support by Fundação para a Ciência e a Tecnologia (FCT) through research grants UIDB/04434/2020 and UIDP/04434/2020 and work contract 2020.03379.CEECIND.
 This research has made use of NASA’s Astrophysics Data System Bibliographic Services.
 
\end{acknowledgements}

%
\bibliographystyle{aa} 
%

\begin{appendix}

\section{MUSE observations}

We used observations from the seven most central fields as these regions are included in the spatial coverage of the \citet{bellini2017a} photometric catalog. A mosaic of these fields is shown in Fig. 1 of \citet{kamann2018} and on our project website\footnote{\url{https://musegc.uni-goettingen.de/index.php/targets?name=ngc5139}}. 
The observations from 2018 and 2019 benefited from the use of the adaptive optic system installed on UT4 of the VLT.
The data reduction was done with the standard MUSE pipeline \citep{weilbacher2020} and the spectral extraction was performed with the \textsc{PampelMuse} software \citep{kamann2013}.
A summary of the observations used for this work is presented in Table \ref{table_obs}.

\begin{table}[h]
\small
\caption{MUSE observations of \omcen}
\label{table_obs}      
\centering                    
\begin{tabular}{c c l c c c}        
\toprule\toprule
\noalign{\vskip4bp}
Field &  RA & DEC & \multicolumn{2}{c}{\# Epochs} & Total exp. time \\
 & & & non-AO & AO & (s) \\
 \midrule
 1 & 13:26:45.0 & $-$47:29:09 & 8 & 5 & 1755 \\
 2 & 13:26:45.0 & $-$47:28:24 & 7 & 5 & 1620 \\
 3 & 13:26:49.5 & $-$47:29:09 & 7  & 6 & 1710 \\
 4 & 13:26:49.5 & $-$47:28:24 & 7  & 6  & 1755 \\
 5 & 13:26:40.6 & $-$47:28:31 & 7  & 6  & 3040 \\
 6 & 13:26:53.1 & $-$47:29:01 & 7  & 6  & 3120 \\
 7 & 13:26:36.8 & $-$47:27:54 & 6  &  6 & 3600 \\
\bottomrule
\end{tabular}
\end{table}

\section{Additional atmospheric properties of the MS stars}\label{app_b}

   \begin{figure}
   \centering
   \includegraphics[width=0.9\columnwidth]{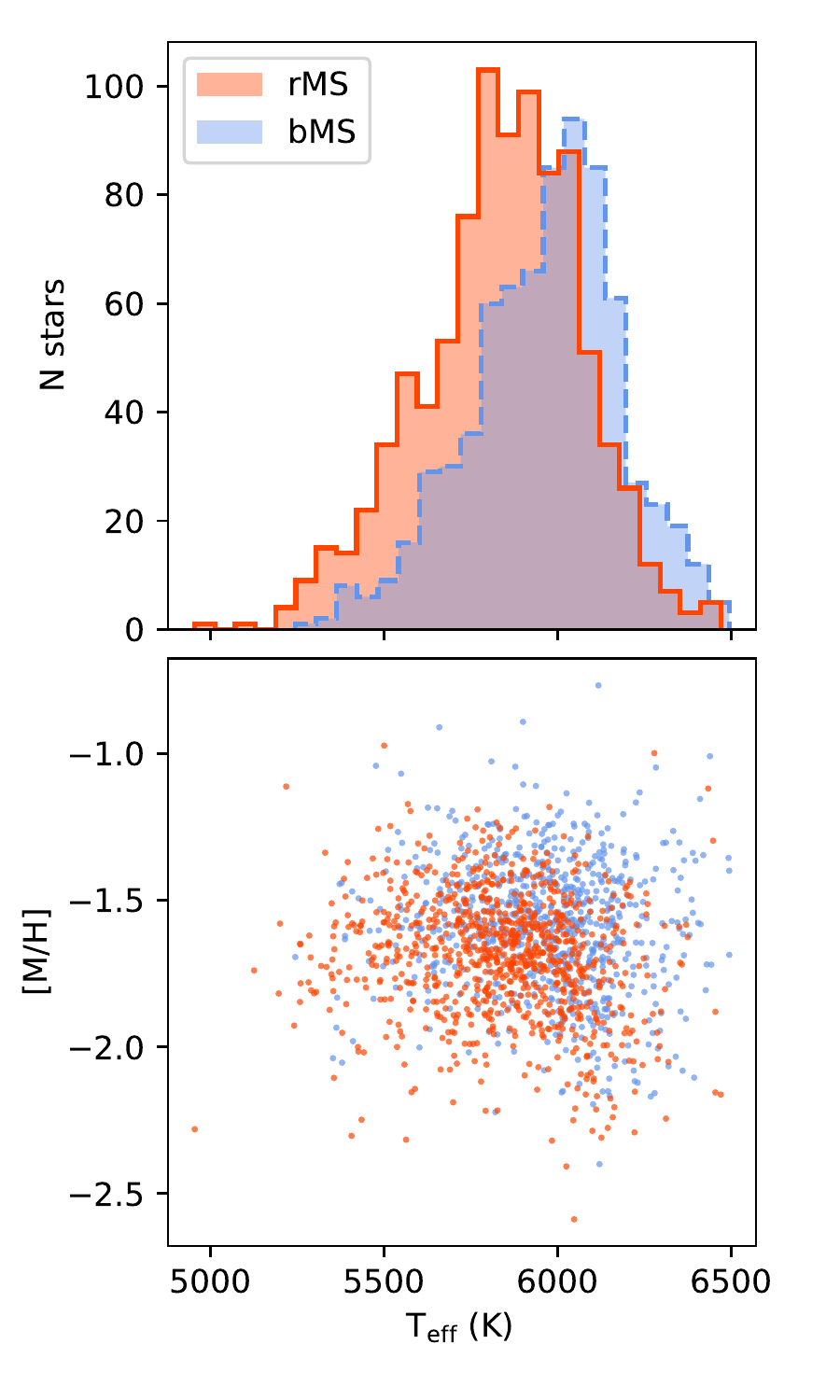}
      \caption{Top: Effective temperature distribution for the rMS and bMS $S/N >$ 20 samples. As expected, the bMS stars are on average hotter than the rMS stars. Bottom: Distribution of the rMS and bMS stars in the \teff$-$\mh\ plane. 
              }
         \label{parameters}
   \end{figure}

The adopted effective temperatures of the stars affect the resulting metallicities. It is stated as the main source of uncertainties on the metallicity difference between the rMS and bMS in PI05 as well as the source of the systematic offset between the $[Fe/H]$ obtained by \citet{marino2011} and previous investigations of RGB stars. As opposed to many spectroscopic investigations of stars in GCs, including that of PI05, our temperatures are not based on the photometric properties of the stars. We used the \teff\ obtained from the isochrones as a starting value for the spectral fitting, but the final temperature was obtained simultaneously with the \mh\ from the spectra themselves. The Balmer lines $H_\beta$ and $H_\alpha$ are the main temperature indicators, while the Mg $b$ and Ca triplet lines are the main metallicity indicators. The median errors on [M/H] and \teff, which were  calibrated as explained in Sect. 2, are $\pm$0.1 dex and 70 K. The fact that the bMS stars are bluer than the rMS in the CMD is an indication that at a given magnitude they are also hotter than their rMS counterparts. We retrieved this general property from our fitting procedure as shown in the top panel of Fig. \ref{parameters}. The median \teff\ for the bMS and rMS stars are 5980 K and 5860 K, respectively. We note here that our MS stars are significantly hotter than the stars analyzed by PI05 who adopted \teff\ of 5400 K and 5200 K for the bMS and rMS, respectively. This difference is mainly due to the fact that our stars are $\approx$0.7 magnitudes brighter than the PI05 sample, thus they are intrinsically hotter. Indeed, our isochrones have differences of $\approx$500 K between an $F606W$ magnitude of 19.7 and 20.5. 
In the bottom panel of Fig. \ref{parameters}, we plotted the rMS and bMS stars (in red and blue, respectively) in the \teff$-$\mh\ plane and the distribution of stars in this diagram does not suggest the presence of a correlation between the two parameters. 
The atmospheric parameters (\teff, log $g$, \mh) and coordinates of the stars included in our rMS, bMS, and in the BE17 sequences are available in Table B.1 which is only available in electronic form at CDS and on our project homepage\footnote{\url{https://musegc.uni-goettingen.de/}}.

\section{Metallicities of the 15 MSs of Bellini et al. 2017}\label{app_c}

\begin{figure}
   \centering
   \includegraphics[width=0.9\columnwidth]{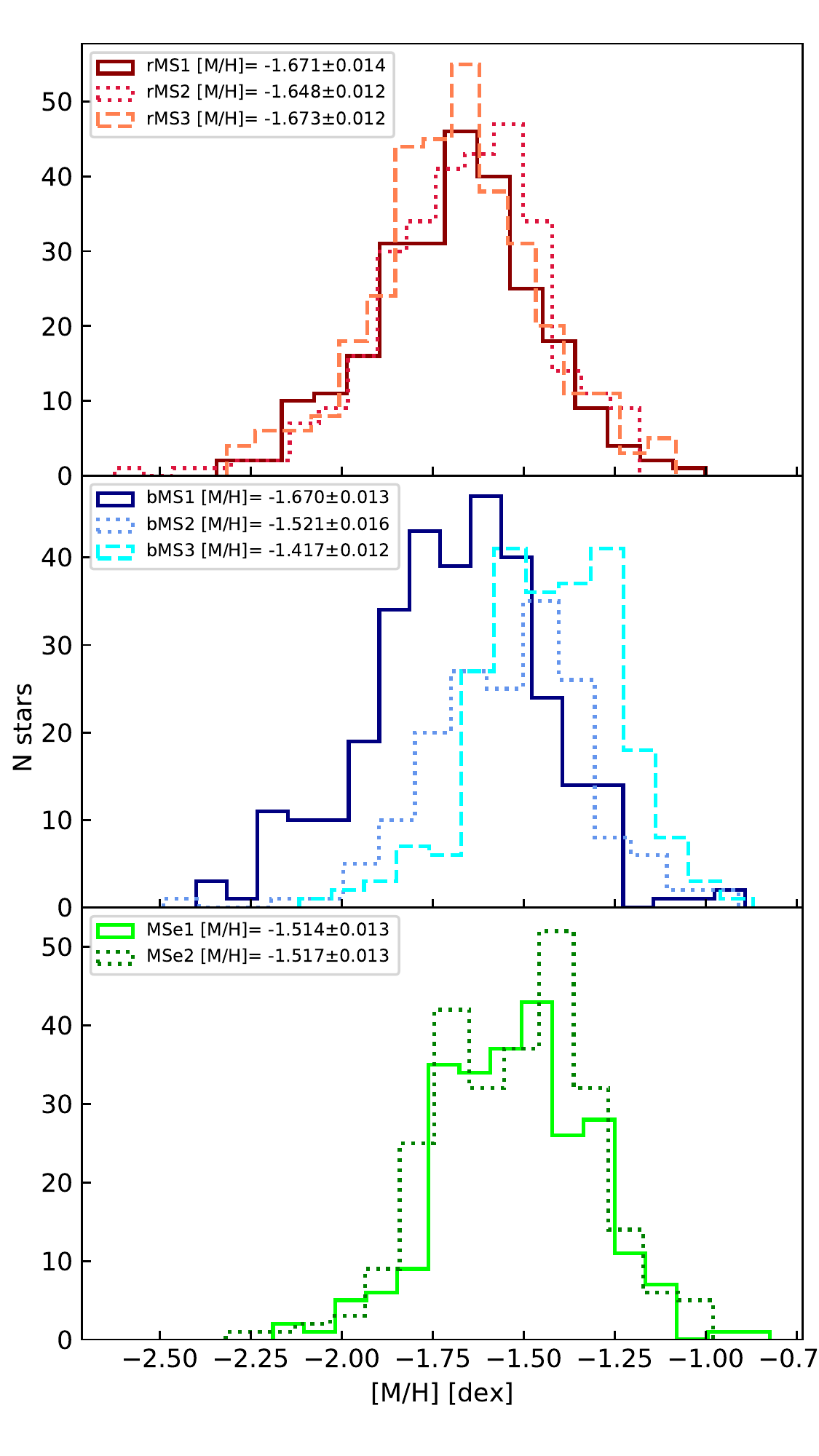}
      \caption{Metallicity distributions of the subpopulations of rMS and bMS as identified by BE17. For the MSe, we plotted the distributions for the two most populous groups: MSe1 and MSe2.
              }
         \label{rMS_bMS_split}
   \end{figure}

\begin{figure*}
   \centering
   \includegraphics[scale=0.95]{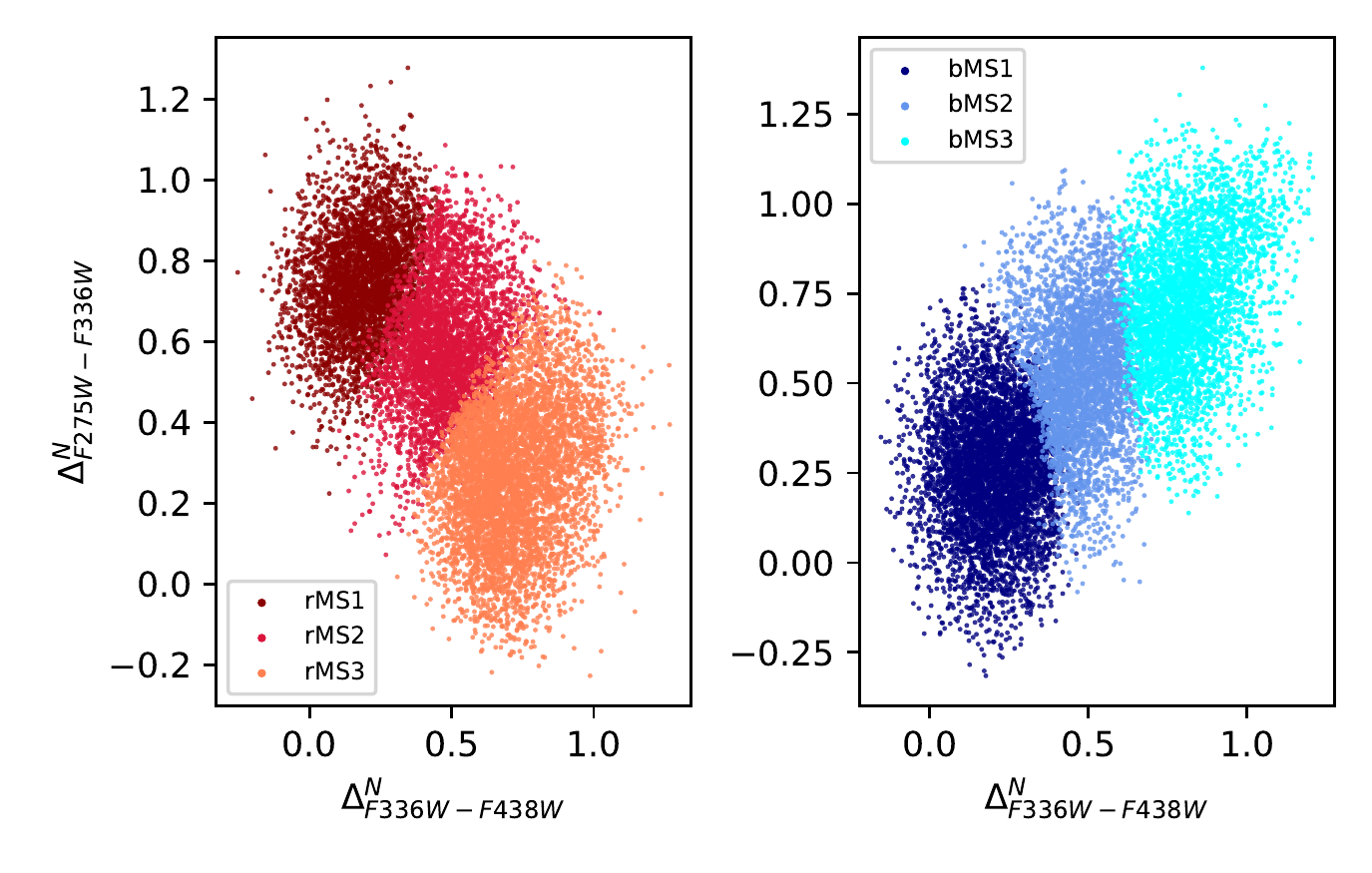}
    \includegraphics[scale=0.9]{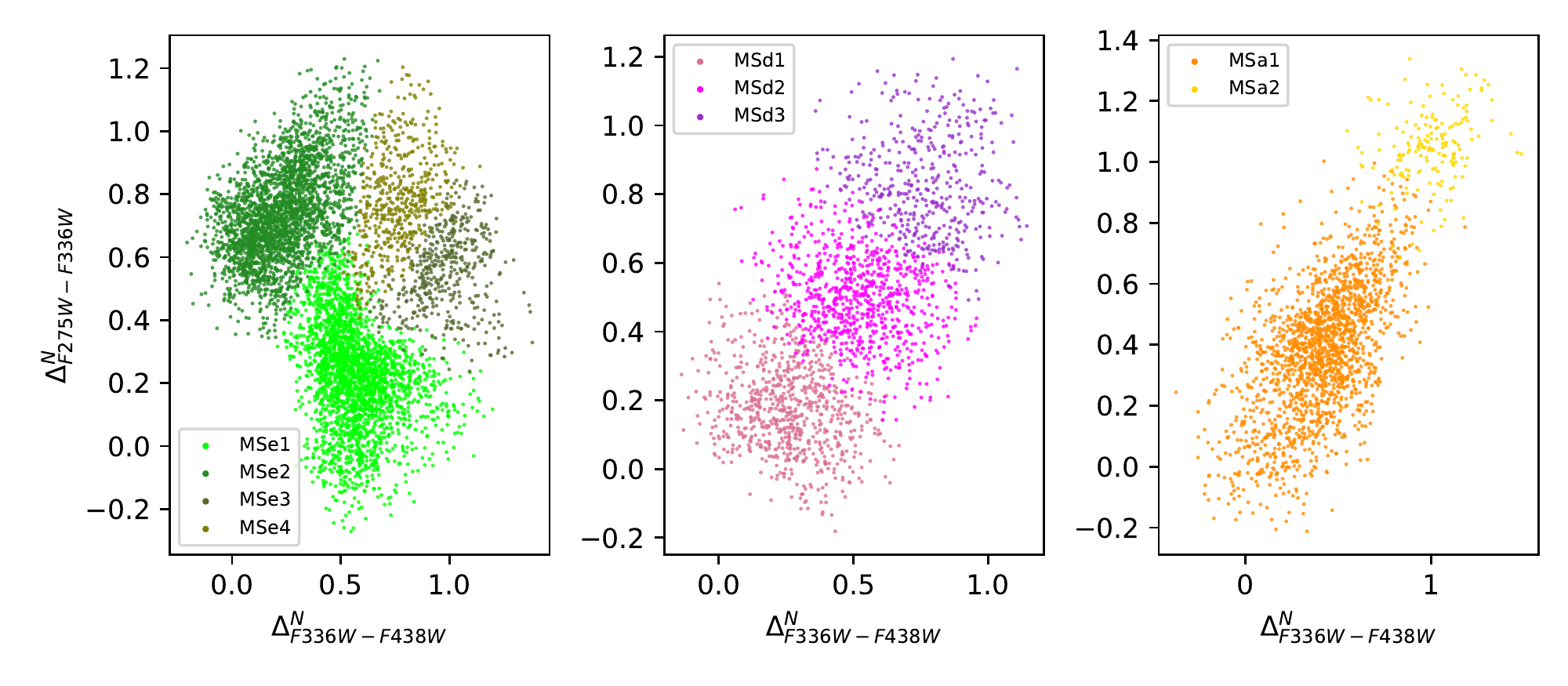}
      \caption{Chromosome maps of the five MSs. The subpopulations of each MS are plotted in different colors. The axis are the two rectified and parallelized pseudo-colors used for the construction of the chromosome maps in BE17.
              }
         \label{Cmap}
   \end{figure*}

For the sake of completeness,  in Table \ref{appendix_table_5MS} we present the properties for the 15 MSs identified in BE17. Unfortunately, some of these subpopulations have very few stars left in our spectroscopic samples. However, the subpopulations of the bMS and rMS are well populated enough to clearly show a different behavior in Fig. \ref{rMS_bMS_split}. As discussed in Sect. 4.2, while the rMS subpopulations share the same metallicity, this is not the case for the bMS. 
It is interesting to note that the bMS and rMS stars also have a different behavior in their respective chromosome map; the bMS stars show populations aligned from the lower left to the upper right, while the rMS stars are aligned from the lower right to the upper left (see Fig.\ref{Cmap}).
It seems possible that variations in iron shape (at least in part), the chromosome map of the bMS stars, and light element variations are responsible for the different disposition of the rMS stars. This is further supported when we look at the result for MSe1 and MSe2, two subpopulations with enough stars to provide a spectroscopic sample of a decent size (see Fig. \ref{rMS_bMS_split}, bottom panel). 
The MSe1 and MSe2 stars have similar metallicities and they are the only other subpopulations that align in the same way as the rMS in the chromosome maps. Based on their photometric properties, BE17 expected the stars of MSe1 to have a similar iron content as the rMS, which is not what we observe. We find the MSe1 stars to be more metal-rich than the rMS by $\approx$0.15 dex. 

There is one last behavior worth mentioning concerning the populations of the bMS, MSd, and MSa. According to BE17, they share similar photometric properties, which is illustrated in their Fig. 15 where the subpopulations are ordered in terms of $\Delta$color (X-$m_{\rm F438W}$ where X is a given filter) in the bluest filters ($\lambda <$ 336 nm) with respect to the rMS1. There is a clear progression from bluer to redder as follows: bMS1, bMS2, bMS3, MSd1, MSd2, MSd3, MSa1, and MSa2.  From our average metallicities, we also see a progression of increasing metallicity in the same order, besides for MSd2 and MSd3 which have the same average \mh\ but also a small spectroscopic sample size. This is consistent with the fact that the bluest filter ($F225W$) is especially sensitive to Fe variations.

\begin{table}
\small
\caption{Properties of the 15 MSs identified in \citet{bellini2017c}}
\label{appendix_table_5MS}      
\centering                    
\begin{tabular}{l c l c c}        
\toprule\toprule
MS &  $S/N$ & \# of stars & Mean $[M/H]$ & $\sigma$ $[M/H]$ \\   
\midrule                      
rMS1 & $>$ 15 & 248 & $-$1.663 $\pm$ 0.013 & 0.174 $\pm$ 0.012 \\
rMS2 & $>$ 15 & 302 & $-$1.641 $\pm$ 0.012 & 0.160 $\pm$ 0.010 \\
rMS3 & $>$ 15 & 329 & $-$1.667 $\pm$ 0.012 & 0.172 $\pm$ 0.011 \\
bMS1 & $>$ 15 & 313 & $-$1.663 $\pm$ 0.013 & 0.187 $\pm$ 0.011 \\
bMS2 & $>$ 15 & 169 & $-$1.517 $\pm$ 0.016 & 0.170 $\pm$ 0.014 \\
bMS3 & $>$ 15 & 231 & $-$1.413 $\pm$ 0.012 & 0.148 $\pm$ 0.011 \\
MSe1 & $>$ 15 & 246 & $-$1.508 $\pm$ 0.013 & 0.161 $\pm$ 0.011 \\
MSe2 & $>$ 15 & 261 & $-$1.510 $\pm$ 0.013 & 0.163 $\pm$ 0.011 \\
MSe3 & $>$ 15 & 27 & $-$1.464 $\pm$ 0.023 & 0.073 $\pm$ 0.034 \\
MSe4 & $>$ 15 & 46 & $-$1.409 $\pm$ 0.028 & 0.148 $\pm$ 0.027 \\
MSe3+4 & $>$ 15 & 73 & $-$1.429 $\pm$ 0.019 & 0.126 $\pm$ 0.019 \\
MSd1 & $>$ 15 & 51 & $-$1.289 $\pm$ 0.021 & 0.115 $\pm$ 0.020 \\
MSd2 & $>$ 15 & 53 & $-$1.178 $\pm$ 0.020 & 0.114 $\pm$ 0.018 \\
MSd3 & $>$ 15 & 45 & $-$1.161 $\pm$ 0.024 & 0.130 $\pm$ 0.022 \\
MSa1 & $>$ 15 & 315 & $-$0.970 $\pm$ 0.009 & 0.140 $\pm$ 0.007 \\
MSa2 & $>$ 15 & 31 & $-$0.862 $\pm$ 0.027 & 0.136 $\pm$ 0.025 \\

\bottomrule                          
\end{tabular}
\end{table}

\end{appendix}

\end{document}